\documentclass[sigconf,natbib]{acmart}

\AtBeginDocument{%
 }

\usepackage{amsmath}
\usepackage{graphicx}
\usepackage{url}
\usepackage{booktabs} 
\usepackage{multirow}
\usepackage{amsfonts}
\usepackage{threeparttable}
\usepackage{placeins}
\usepackage{subfigure}
\usepackage{algorithmic}
\usepackage{textcomp}
\usepackage{xcolor}
\usepackage{colortbl}

\usepackage{color}

\copyrightyear{2023}
\acmYear{2023}
\setcopyright{rightsretained}
\acmConference[SIGIR '23]{Proceedings of the 46th International ACM SIGIR Conference on Research and Development in Information Retrieval}{July 23--27, 2023}{Taipei, Taiwan}
\acmBooktitle{Proceedings of the 46th International ACM SIGIR Conference on Research and Development in Information Retrieval (SIGIR '23), July 23--27, 2023, Taipei, Taiwan}
\acmDOI{10.1145/3539618.3592021} 
\acmISBN{978-1-4503-9408-6/23/07}

\makeatletter
\gdef\@copyrightpermission{
  \begin{minipage}{0.3\columnwidth}
   \href{https://creativecommons.org/licenses/by/4.0/}{\includegraphics[width=0.90\textwidth]{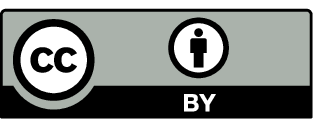}}
  \end{minipage}\hfill
  \begin{minipage}{0.7\columnwidth}
   \href{https://creativecommons.org/licenses/by/4.0/}{This work is licensed under a Creative Commons Attribution International 4.0 License.}
  \end{minipage}
  \vspace{5pt}
}
\makeatother

\begin{document}

\title{Mining Interest Trends and Adaptively Assigning Sample Weight for Session-based Recommendation}

\author{Kai Ouyang}
\authornote{Both authors contributed equally to this research.}
\email{oyk20@mails.tsinghua.edu.cn}
\orcid{0000-0002-0884-529X}
\author{Xianghong Xu}
\authornotemark[1]
\email{xxh20@mails.tsinghua.edu.cn}
\orcid{0000-0003-2447-4107}
\affiliation{%
  \institution{SIGS, Tsinghua University}
  \city{Shenzhen}
  \country{China}
  \postcode{518055}
}


\author{Miaoxin Chen}
\email{cmx20@mails.tsinghua.edu.cn}
\orcid{0000-0003-3518-9555}
\affiliation{%
  \institution{Shezhen International Graduate School, Tsinghua University}
  \city{Shenzhen}
  \country{China}
  \postcode{518055}
}

\author{Zuotong Xie}
\email{xiezt20@mails.tsinghua.edu.cn}
\orcid{0009-0009-2362-0214}
\affiliation{%
  \institution{Shezhen International Graduate School, Tsinghua University}
  \city{Shenzhen}
  \country{China}
  \postcode{518055}
}

\author{Hai-Tao Zheng}
\authornote{Corresponding author.}
\email{zheng.haitao@sz.tsinghua.edu.cn}
\orcid{0000-0001-5128-5649}
\affiliation{%
  \institution{Shezhen International Graduate School, Tsinghua University}
  \city{Shenzhen}
  \country{China}
  \postcode{518055}
}

\author{Shuangyong Song}
\email{songshy@chinatelecom.cn}
\orcid{0000-0001-7465-1082}
\affiliation{%
  \institution{China Telecom Corporation Ltd. Data\&AI Technology Company}
  \city{Beijing}
  \country{China}
}

\author{Yu Zhao}
\email{zhaoy11@chinatelecom.cn}
\orcid{0009-0007-6985-6105}
\affiliation{%
  \institution{China Telecom Corporation Ltd. Data\&AI Technology Company}
  \city{Beijing}
  \country{China}
}








\renewcommand{\shortauthors}{Kai Ouyang et al.}

\begin{abstract}
Session-based Recommendation (SR) aims to predict users' next click based on their behavior within a short period, which is crucial for online platforms. 
However, most existing SR methods somewhat ignore the fact that user preference is not necessarily strongly related to the order of interactions.
Moreover, they ignore the differences in importance between different samples, which limits the model-fitting performance.
To tackle these issues, we put forward the method, \textbf{M}ining Interest \textbf{T}rends and \textbf{A}daptively Assigning Sample \textbf{W}eight, abbreviated as \textbf{MTAW}. 
Specifically, we model users' instant interest based on their present behavior and all their previous behaviors. 
Meanwhile, we discriminatively integrate instant interests to capture the changing trend of user interest to make more personalized recommendations.
Furthermore, we devise a novel loss function that dynamically weights the samples according to their prediction difficulty in the current epoch. 
Extensive experimental results on two benchmark datasets demonstrate the effectiveness and superiority of our method.

\end{abstract}




\maketitle

\section{Introduction}
\label{sec:introduction}
In many real-world recommendation scenarios, users are usually unable to be identified or tracked due to privacy policies. Therefore, recommendation systems need to discern latent user interests based on sparse interaction data in the absence of user information. To address this challenge, Session-based Recommendation (SR) has been developed. With the boom of e-commerce platforms, SR has attracted increasing attention from academia and industry.

Most existing research efforts in SR regard the session as a strictly ordered sequence~\cite{acrec_xu2022modeling}. 
At the beginning of SR research, many RNN-based models~\cite{gru4rec_hidasi2015session, rnn1_jannach2017recurrent, rnn2_hidasi2018recurrent, rnn2_tan2016improved} have been proposed. 
For example, GRU4Rec~\cite{gru4rec_hidasi2015session} captures user latent interest by modeling the sequence information of user interactions.
In recent years, researchers have proposed many GNN-based SR models~\cite{srgnn_wu2019session, GCSAN_xu2019graph, gce_wang2020global, gnnsr2_qiu2020exploiting, fgnn_qiu2019rethinking, ouyang-etal-2022-social}. 
They model the session as a directed graph and focus on modeling the transitions between adjacent interactions with the pairwise relationships of nodes. 
Besides, most SR methods treat all samples equally during model training. 
They are optimized by treating samples indiscriminately with cross-entropy or InfoNCE~\cite{InfoNCE_oord2018representation} loss function. 
These methods achieve considerable performance improvement due to their effective capture of sequence information and optimization way like contrastive learning.

However, they have two defects.
\textbf{(a)} These SR models implicitly follow a strong assumption, \emph{i.e.}, the relative order of adjacent interactions is strongly  associated with the users' interests.
However, the relative order of users' behaviors does not have an absolute correlation with their interests, and some behaviors may just be noise signals.
For instance, some user interactions may result from accidental clicking or the random recommendation function of APPs. 
Thus, strictly modeling timing can increase the impact of these noises, ultimately limiting the model's performance. 
\textbf{(b)} They ignore the difference between samples. 
In reality, different sessions have varying numbers and credibility of interactions, resulting in different prediction difficulties.  
This, in turn, makes their importance in model fitting different.  
Hence, treating all samples equally reduces the fitting performance of the model.

To tackle these issues, we propose  capturing the changing trend of user interest, rather than focusing on modeling the relative order between interactions.
Moreover, we suggest assigning importance weight to the samples, rather than treating them equally.
Hence, we put forward method, \textbf{M}ining Interest \textbf{T}rends and \textbf{A}daptively Assigning Sample \textbf{W}eight, abbreviated as \textbf{MTAW}.
Specifically, we capture the instant interest at the current moment according to user interactions at the previous moments and their position information.
We integrate these instant interests discriminatively to mine the changing trend of user interest. 
Compared to order information, trend information on interest changing is more reliable. 
Additionally, we devise the Adaptive Weight (AW) loss function, which adaptively assigns weights to different samples according to their prediction difficulty in the current epoch. 
This makes the model pay more attention to hard samples in the model-fitting process, improving its effectiveness.
To summarize, we make the following contributions:
\begin{itemize}
    \item We model the changing trends of users' interests. Technically, we track users' instant interests at each moment and then integrate these interests discriminatively to achieve more personalized recommendations.
    \item We devise the AW loss function. It adaptively assigns weights to different samples and enables us to focus more on difficult samples to enhance the model-fitting effect.
    \item Experimental results on two datasets demonstrate that MTAW overwhelmingly outperforms the state-of-the-art (SOTA) methods. Moreover,  MTAW is far more efficient and requires fewer parameters than SOTA methods.
\end{itemize}

\section{Method}

\begin{figure}[t]
\centering
\small	
\includegraphics[width=0.48\textwidth]{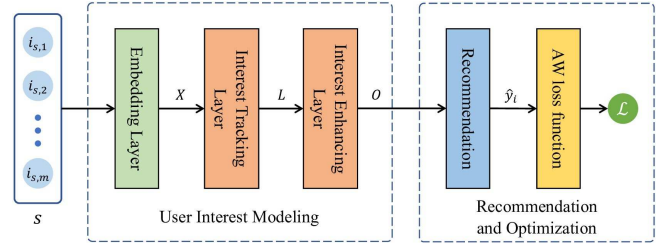}
\caption{The architecture of MTAW.}
\label{fig:arch}
\end{figure}

The basic idea of MTAW is to mine the trend of user dynamic interest and assign different importance weights to different samples.
The architecture of MTAW is depicted in Figure~\ref{fig:arch}. It can be mainly divided into the following parts: 
\textbf{(1)} User Interest Modeling, which takes two steps to capture the user's interest evolving: \textit{(i)} Interest Tracking Layer.
\textit{(ii)} Interest Enhancing Layer.
\textbf{(2)} Recommendation and Optimization, which makes recommendations based on the session representation and assigns different weights to samples according to their difficulty.

\subsection{Problem Statement}

Let $I=\{i_1,i_2,\ldots, i_N\}$ denote the set of all unique items involved in all the sessions, where $N$ is the total number of items. Each session is represented as a list $s=[i_{s,1},i_{s,2},\ldots,i_{s,m}]$ ordered by timestamps, where $m$ is the length of the session $s$ and $i_{s,k}\in I (1\leq k\leq m)$ represents a clicked item of the user within the session $s$. The goal of the task is to predict the next click for the session $s$, \emph{i.e.}, the sequence label $i_{s,m+1}$. For the session $s$, we calculate the probabilities $\hat{y}$ for all possible items, where the recommendation score of an item is the corresponding element of vector $\hat{y}$.
The items corresponding to the top-K scores will be recommended.

\subsection{User Interest Modeling}
To convert the input session into vectors, we construct the Embedding Layer.
For each item $i$ in the input session, the hidden representation is:
\begin{equation}
    x_i = e_i + p_i, \label{embedding:sr}
\end{equation}
where $e_i \in \mathbb{R}^{d}$ is item embedding, $d$ is the embedding size, $p_i \in \mathbb{R}^{d}$ is the position embedding, and $x_i \in \mathbb{R}^{d}$ denotes the hidden representation of item $i$.
Besides, we use $X = \{x_1,x_2, \dots, x_m\}$ to denote the embedding set of session $s=[i_{1},i_{2},\ldots,i_{m}]$.

\subsubsection{Interest Tracking Layer}
User interests are usually dynamic, and user behavior sequence is the carrier of user interest. 
Therefore, to capture user real interest, we need to mine the instant interest sequence based on the user behavior sequence.
Specifically, we employ the Attention network to extract user instant interest at the current moment according to its previous interactions.
Formally, the Attention network can be defined as:
\begin{equation}
\mathrm{Attention}(Q, K, V) = \mathrm{softmax}(\frac{Q^{\top}K}{\sqrt{d}})V,
\end{equation}
where $Q, K$, and $V$ are the input matrices.
To ensure that the extracts for $t$-th item can depend only on its previous items, we derive $m$ slices from $X$ in chronological order: $X'_1=\{x_1\}$, $X'_2=\{x_1, x_2\}$, $\ldots$, $X'_m=\{x_1,x_2,\ldots,x_m\}$. 
Then, for each slice like $X'_t=\{x_1,x_2,\ldots,x_t\}$, where $t \leq m$ represents the serial number of slices, we adopt the Attention network to extract user instant interest:
\begin{equation}
\begin{split}
    &Q_x=\mathrm{ReLU}(\mathrm{MLP}(x_t)), \\
    &l'_t=\mathrm{Attention}(Q_x, X'_t, X'_t), \\
\end{split}
\end{equation}
where $l'_t \in \mathbb{R}^{d}$, MLP denotes the multi-layer perceptron, and ReLU is the activation function. 
We perform the above process in parallel by using a mask matrix in the Attention network.
The output is hidden state set $L'=\{l'_1, l'_2, \ldots, l'_m\}$ for session $s$.
Moreover, we apply the \textit{Position-wise Feed-Forward Network (FFN)} to endow the model with more non-linearity:
\begin{equation}
    L = FFN(L') = \mathrm{MLP}(\mathrm{ReLU}(\mathrm{MLP}(L'))),
\end{equation}
where two MLPs represent two different multi-layer perceptrons. 
Then, we add a residual connection and layer normalization on the result to alleviate the instability of the model training. 
We also add the dropout mechanism to alleviate the overfitting. 
For simplicity, we denote the \textit{Interest Tracking Layer} as ITL, \emph{i.e.},
    $L=\mathrm{ITL}(X)$,
where $L=\{l_1, l_2, \ldots, l_m\},\ l_m \in \mathbb{R}^{d}$ is the final output of this layer, and each of them denotes the user instant interest of current interaction.

\subsubsection{Interest Enhancing Layer}
The evolution of user interest will directly affect the user's choice of the next item. Therefore, we design the Interest Enhancing Layer to conduct in-depth excavation and analysis of the evolution process of the user's interest. 
Besides, this layer injects information about interest evolution trends into the session representation. 

Specifically, to obtain the changing trend of users' interest in the next item, we adopt the same Attention network to discriminatively integrate the instant interests:
\begin{equation}
    O = \mathrm{Attention}(Q_l, L, L),
\end{equation}
where $Q_l$ is the last element $l_m$ of $L$, and $O \in \mathbb{R}^{d}$ represent the session representation. 
In a word, we use the user's last instant interest to be the query, and the whole instant interest sequence to be the key. 
It can learn the changing trend that hides behind the evolution process of user interest.

\subsection{Recommendation and Optimization}
In this layer, we complete the prediction of the current session and the model optimization process.

\textit{Recommendation.} To make recommendations, for each item $i \in I$, we get the ranking score as follows:
\begin{equation}
\begin{split}
    &\hat{O} = \mathrm{L2Norm}(O),\ \hat{x}_i = \mathrm{L2Norm}(x_i), \\
    &\hat{y}_i = \mathrm{softmax}(\hat{O}^{T}\hat{x}_i), \\
\end{split}
\end{equation}
where $x_i$ is the embedding of item $i$, L2Norm is the L2 Normalization function and $\hat{y_i}$ denotes the final probability of item $i$.

\textit{Optimization.} There are differences between samples and the difficulty of models in predicting different sessions.
Besides, easily classified negatives comprise the majority of the loss and dominate the gradient \cite{focal_lin2017focal}.
Therefore, we propose to assign different weights to different samples.
Inspired by Focal loss \cite{focal_lin2017focal}, we assign weights based on the prediction deviation of samples in the current epoch. 

More formally, we propose the Adaptive Weight (AW) loss function which adds a modulating factor to the cross-entropy loss function.
The AW Loss can be formulated as follows:

\begin{equation}
\begin{split}
    &p_{i}=
\left\{
\begin{array}{rcl}
\hat{y}_i,      &      & \mathrm{if}\ y = 1, \\
1 - \hat{y}_i,      &      & \mathrm{otherwise}, \\
\end{array} 
\right. \\
& \mathcal{L} = - \sum_{i=1}^{M}  (2-2p_i)^{\gamma}\log(p_i),
\end{split}
\label{eq2}
\end{equation}
where $\gamma$ is the temperature coefficient and as $\gamma$ is increased the effect of the modulating factor is increased.
$y$ is the ground truth probability distribution of the next item, which is a one-hot vector.
$M$ is the total number of samples.
Since $p_i \in [0, 1]$, $(2-2p_i)$ is around 1.
$(2-2p_i)^{\gamma}$  indicates the deviation between the predicted value and the ground truth of the sample, \emph{i.e.}, the difficulty of the sample in the current epoch.
Intuitively, the modulating factor reduces the loss contribution of simple samples and expands that of hard samples.
\section{Experiments}

\subsection{Settings}
\subsubsection{Datasets and Metrics.}
For a fair comparison, we conduct the experiment on the two public datasets: \textbf{Tmall}\footnote{https://tianchi.aliyun.com/dataset/dataDetail?dataId=42} comes from IJCAI-15 competition, which contains anonymous shopping logs on Tmall online platform.
\textbf{RetailRocket}\footnote{https://www.kaggle.com/retailrocket/ecommerce-dataset} comes from a Kaggle contest and contains the browsing activity of users within six months.
For a fair comparison, we implement our model on the public pre-processed version datasets provided by $S^{2}$-DHCN~\footnote{https://github.com/xiaxin1998/DHCN}. 
The statistics of the two datasets after preprocessing are represented in Table \ref{tb:dataset}.

We use the same evaluation metrics as the previous works \cite{s2dhcn_xia2020self,srgnn_wu2019session}: MRR@K (Mean Reciprocal Rank at K) and P@K (Precision at K). The values of K include 10 and 20. 

\begin{table}[ht]
	\caption{Statistics of RetailRocket and Tmall Datasets}\label{tb:dataset}
	\vspace{-5pt}
    \centering
    \resizebox{0.44\textwidth}{!}{
		\begin{tabular}{cccccc}
			\toprule
			Dataset & \# training  & \# test& \# items & Avg. Len.\\ 
			\midrule
			 RetailRocket &433,643 & 15,132 & 36,968& 5.43 \\
			 Tmall &  351,268 &  25,898 & 40,728 & 6.69 \\
			\bottomrule
		\end{tabular}
	}
\end{table}

\begin{table*}[h]
	\caption{Performance comparison on two datasets (\%). In each metric, the best result is highlighted in boldface and the second best is underlined. And $\dagger$ indicates statistic significant improvement over all baseline models for t-test with $p$-value $<$ 0.01. }
	
	\begin{center}
		{
	\resizebox{0.8\textwidth}{!}{
		\begin{tabular}{ccccccccc}
			\toprule
			\multirow{2}{*}{Method} &
			 \multicolumn{4}{c}{Tmall} & \multicolumn{4}{c}{RetailRocket} \cr
			\cmidrule(lr){2-5}\cmidrule(lr){6-9} & P@10 & MRR@10 & P@20 & MRR@20 & P@10 & MRR@10 & P@20 & MRR@20  \\ 
			\midrule

			FPMC \cite{fpmc_rendle2010factorizing}  &13.10& 7.12 &16.06 &7.32 & 25.99 & 13.38 & 32.37 & 13.82 \\
				
			GRU4REC \cite{gru4rec_hidasi2015session}  &14.16 & 6.56 & 18.20 & 6.85 & 34.41 & 15.06 & 44.89 & 15.77  \\
			
			NARM \cite{narm_li2017neural}   &19.17 &10.42& 23.30 &10.70 & 42.07  & 24.88 & 50.22 & 24.59 \\
			
			STAMP \cite{stamp_liu2018stamp}   &22.63 &13.12 &26.47 &13.36 & 42.95 & 24.61 & 50.96 & 25.17 \\
			
			SASRec \cite{sasrec_kang2018self} &  22.06 &14.02 &26.95 &14.21 & 44.65 & 25.53 & 51.12 & 25.91 \\
			
			NextItNet \cite{cnnrec_yuan2019simple} & 22.67 & 13.12 & 27.22 & 13.32 &41.12 &23.99 &48.26 & 24.48   \\
			
			SR-GNN \cite{srgnn_wu2019session}  &23.41 &13.45 &27.57& 13.72 & 43.21 & 26.07 & 50.32 & 26.57   \\
			
			GC-SAN \cite{GCSAN_xu2019graph} &21.32 &12.43 & 25.38 & 12.72 & 43.21 & 26.07 & 50.32 & 26.57   \\
			
			GCE-GNN \cite{gce_wang2020global}  & \underline{28.02} & \underline{15.08} & \underline{33.42} & \underline{15.42} &46.05 &\underline{27.48} &53.63 & \underline{28.01}  \\

			$S^{2}$-DHCN \cite{s2dhcn_xia2020self}  & 26.22 & 14.60 & 31.42 & 15.05 & \underline{46.15} &  {26.85}& \underline{53.66} & {27.30} \\ 
			
			\midrule
			
			MTAW & \textbf{31.67}$^\dagger$ & \textbf{18.90}$^\dagger$ & \textbf{37.17}$^\dagger$ & \textbf{19.14}$^\dagger$ & \textbf{48.41}$^\dagger$ & \textbf{29.96}$^\dagger$ & \textbf{56.39}$^\dagger$ & \textbf{30.52}$^\dagger$ \\ 
			Improv. (\%)  & 13.03 & 25.33 & 11.22 & 24.12 & 4.90 & 9.02 & 5.09 & 8.96 \\
			\bottomrule
		\end{tabular}}
		}
	\end{center}
	\label{tb:overall_result}
\end{table*}

\subsubsection{Baselines and Implementation Details.}
We compare our model with the following representative SR methods:
\textbf{FPMC} \cite{fpmc_rendle2010factorizing} is a sequential method based on Markov Chain. 
\textbf{GRU4REC} \cite{gru4rec_hidasi2015session}, which is a representative sequential model based on Gated Recurrent Unit (GRU).
\textbf{NARM} \cite{narm_li2017neural} utilizes the self-attention mechanism and GRU to capture the main purpose of the session.
\textbf{STAMP} \cite{stamp_liu2018stamp}, which uses the self-attention mechanism to represent the intent of the session, and emphasizes the importance of the last click in each session.
\textbf{SASRec} \cite{sasrec_kang2018self} solely uses self-attention mechanism.
\textbf{NextItNet} \cite{cnnrec_yuan2019simple}, which is the best performing CNN-based SBR model. 
\textbf{SR-GNN} \cite{srgnn_wu2019session} models separated session sequences into graph-structured data and uses graph neural networks to capture complex item transitions.
\textbf{GC-SAN} \cite{GCSAN_xu2019graph}, which combines GNN and multi-layer self-attention to make recommendations.
\textbf{GCE-GNN} \cite{gce_wang2020global} is a widely compared GNN-based SBR model that learns global and local information of sessions.
\textbf{$S^{2}$-DHCN} \cite{s2dhcn_xia2020self} constructs two types of hypergraphs to learn inter- and intra-session information and introduces self-supervised learning.

For a fair comparison, we follow $S^{2}$-DHCN to make the following settings:
We use Adam~\cite{adam_kingma2014adam} optimizer with a learning rate of 0.001.
We set the embedding size to 100, the number of epochs to 50, and the batch size to 100. 
Besides, we set $\gamma$ to 2 for the Tmall dataset, and set $\gamma$ to 6 for the RetailRocket dataset.

\subsection{Results}
\subsubsection{Overall Performance.}
The experimental results of all models are reported in Table \ref{tb:overall_result}. Based on the results, we can draw the following conclusions: 
\textbf{(a)} These traditional models (\emph{e.g.}, Item-KNN, FPMC) can even outperform the first method based on RNNs (\emph{i.e.}, GRU4REC) in terms of some metrics. It indicates that solely modeling a session as a strictly ordered sequence may result in limiting the ability to capture users' genuine interest. 
GNNs-based models outperform most models because GNNs can model the transitions between adjacent items. However, they model the session as a directed graph, which actually models the session as a strictly ordered sequence, so their performance is surpassed by MTAW. 
\textbf{(b)} MTAW overwhelmingly outperforms all baseline models. This demonstrates the superiority of interest trend modeling and adaptive assignment of sample weights.

\begin{figure}[ht]
\centering
\small	
\includegraphics[width=0.36\textwidth]{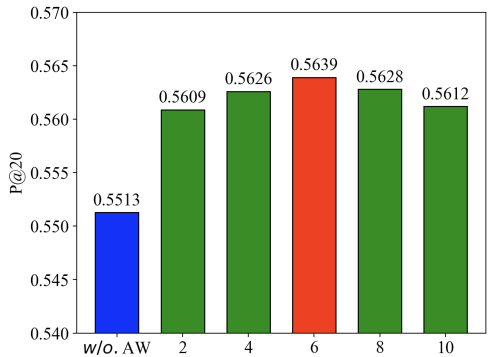}
\caption{AW Loss Study on RetailRocket.}
\vspace{-5pt}
\label{fig:awl}
\end{figure}

\subsubsection{Study on AW Loss Function.} 
To verify the superiority of the AW loss function, and investigate the impact of hyper-parameter $\gamma$ on the final performance, we conduct further experiments.
We search the $\gamma$ in the range of [2, 4, \ldots, 10] in terms of P@20 on the RetailRocket dataset.
Besides, we compare them with variant $w/o.$ AW which uses the normal cross-entropy loss function for optimization.
The results are shown in Figure \ref{fig:awl}, and we can conclude that with the increase of the value $\gamma$, the performance of MTAW first increases and then decreases.
When the $\gamma$ is set to 6, MTAW performs best.
Moreover, when MTAW is optimized without the AW loss function, the performance will decline, which shows the effectiveness of the AW loss function.

\begin{table}[ht]
\vspace{-5pt}
    \centering
    \caption{Training time per epoch and the number of trainable parameters, where s, m, and M respectively represent second, minute, and million.}
    \label{tb:efficiency}
	\resizebox{0.4\textwidth}{!}{
    \begin{threeparttable}{
		\begin{tabular}{cccccc}
			\toprule
		\multirow{2}{*}{Method} &
				 \multicolumn{2}{c}{Tmall} & \multicolumn{2}{c}{RetailRocket} \cr
				\cmidrule(lr){2-3}\cmidrule(lr){4-5}  & Time & \#Params & Time & \#Params   \\ 
			\midrule
			 NextItNet & 34m51s  & 4.23M & 64m27s & 3.85M  \\
			 SR-GNN &4m7s &4.23M & 23m54s &3.86M \\
			 GC-SAN & 3m42s&4.24M &12m20s & 3.87M \\
			 GCE-GNN &2m20s  & 4.35M & 16m02s & 3.98M \\
			 $S^{2}$-DHCN &32m34s & 4.31M &77m12m & 3.94M \\
			 \midrule
			 MTAW   & \textbf{47s} & \textbf{4.02M} &\textbf{58s} & \textbf{3.69M} \\
			\bottomrule
		\end{tabular}}
		
    \end{threeparttable}
	}
 \vspace{-5pt}
\end{table}

\subsubsection{Efficiency Comparison.}
To evaluate the efficiency of MTAW, we compare the training time per epoch and trainable parameters with recent SOTA models on the same device. 
The results are shown in Table \ref{tb:efficiency}.
We can observe that MTAW is far more efficient and requires fewer parameters than recent SOTA methods. 
Compared with GCE-GNN \cite{gce_wang2020global}, MTAW achieves 16.59$\times$ speed up  with fewer parameters on the RetailRocket dataset.
We can conclude that MTAW is both efficient and effective, and modeling interest trends is a potential future work.

\section{Conclusion}
In this paper, we propose MTAW, which mines the changing trend of user interest and adaptively adjusts sample weights for SR.
In MTAW, we use the attention mechanism to capture users' instantaneous interests. Furthermore, we integrate these interests to mine the trend of changing interests.
Additionally, we devise the AW loss function to dynamically assign sample weights.
Extensive experiments on two datasets demonstrate the superiority of MTAW.

\section*{Acknowledgment}
This research is supported by National Natural Science Foundation of China (Grant No.62276154), Research  Center for Computer Network (Shenzhen) Ministry of Education, Beijing Academy of Artificial Intelligence (BAAI), the Natural Science Foundation of Guangdong Province (Grant No.  2023A1515012914), Basic Research Fund of Shenzhen City (Grant No. JCYJ20210324120012033 and JSGG20210802154402007), and the Major Key Project of PCL for Experiments and Applications (PCL2021A06).




\bibliographystyle{ACM-Reference-Format}
\bibliography{sample-sigconf}










\end{document}